\documentclass[sn-mathphys]{sn-jnl}

\jyear{2021}%

\theoremstyle{thmstyleone}%
\theoremstyle{thmstyletwo}%

\theoremstyle{thmstylethree}%

\begin{document}
\title[Article Title]{Joint Antenna Selection and Power Allocation in Massive MIMO Systems with Cell Division Technique for MRT and ZF Precoding Schemes}


\author[]{\fnm{Abdolrasoul} \sur{Sakhaei Gharagezlou}}\email{a.sakhaei@tabrizu.ac.ir}
\author[]{\fnm{Nima} \sur{Imani}}\email{nimaimani91@gmail.com}
\author*[]{\fnm{Mahdi} \sur{Nangir}}\email{nangir@tabrizu.ac.ir}

\affil[]{\orgdiv{Department of Electrical and Computer Engineering}, \orgname{University of Tabriz}, \orgaddress{\street{29 Bahman Blvd}, \city{Tabriz}, \state{East Azarbijan}, \country{Iran}}}


\abstract{One of the most important challenges in the fifth generation (5G) of telecommunication systems is the efficiency of energy and spectrum. Massive multiple-input multiple-output (MIMO) systems have been proposed by researchers to resolve existing challenges. In the proposed system model of this paper, there is a base station (BS) around which several users and an eavesdropper (EVA) are evenly distributed. The information transmitted between BS and users is disrupted by an EVA, which highlights the importance of secure transfer. This paper analyzes secure energy efficiency (EE) of a massive MIMO system, and its purpose is to maximize the secure EE of the system. Several scenarios are considered to evaluate achieving the desired goal. To maximize the secure EE, selecting optimal number of antennas and cell division methods are employed. Each of these two methods is applied in a system with the maximum ratio transmission (MRT) and the zero forcing (ZF) precodings, and then the problem is solved. Maximum transmission power and minimum secure rate for users insert limitations to the optimization problem. Channel state information (CSI) is generally imperfect for users in any method, while CSI of the EVA is considered perfect as the worst case. Four iterative algorithms are designed to provide numerical assessments. The first algorithm calculates the optimal power of users without utilizing existing methods, the second one is related to the cell division method, the third one is based on the strategy of selecting optimal number of antennas, and forth one is based on a hybrid strategy.}




\maketitle

\section{Introduction}\label{sec1}
Telecommunication systems have always had challenges, such as the need for high speed data communication, video broadcasting requirements, stable and reliable transmission on the movie, and so on. These challenges mainly stem from increasing the number of users in telecommunication systems. On the other hand, there is a need to increase spectral efficiency (SE) and energy efficiency (EE), which have been addressed by developing telecommunication systems to meet them. One of these developments is multiple-input multiple-output (MIMO) systems, which can provide high SE and reliability. MIMO systems confront the problem of complexity in relationships and computations. To solve this problem, linear receivers and transmitters are considerably used for linearizing relationships [1]-[2].

The performance of MIMO systems is ameliorated by using a large number of antennas. If the base station (BS) connects to hundreds of users with hundreds of antennas simultaneously in a same frequency-time block, then it is called a massive MIMO system. These systems use antenna arrays with more than one hundred elements for direct energy transferring. Specifically, by using $M$ antennas at the BS for $K$ single antenna users, a multiplexing with degree of $M$ can be achieved. For instance, assuming that the number of mobile stations is constant, by doubling the number of antennas of the BS, the transmitted power is reduced by half and the SE remains constant; while the EE is doubled in this case [3].

The total users' rate optimization problem for the MIMO-NOMA system is inquired in [4]. The problem is constrained by the total transmission power and the minimum rate value to meet the quality of service (QoS), which is required by users. In [4], the channel state information (CSI) conditions and non-orthogonal multiple access (NOMA) MIMO system model, which can reach the full transmission rate, are examined. Full transmission rate means that the submission rate of users is equal to the channel capacity of them. The optimal power allocation method is proposed using the CSI. The obtained results show that the proposed method performs better than the traditional single-user and multi-user time division multiple access method.

In [5], to maintain the confidential information of massive NOMA-MIMO networks, the use of artificial noise (AN) is emphasized. The BS encrypts data and injects AN based on the CSI estimation. Next, a secure ergodic transfer rate is obtained in the downlink. Confidential performance analysis to consider the effect of key parameters is also performed sequentially for a large number of transmitter antennas and total transmit power at the BS. In order to maximize the confidential rate of the system based on the obtained ergodic rate, it is proposed to allocate power jointly to the training phase of the uplink and the transfer phase of the downlink. Further, another optimization algorithm is proposed to maximize the secure EE. The simulation results demonstrate that NOMA networks with massive MIMO and AN method, greatly benefit to the system in terms of the confidential performance.

In massive MIMO systems, one of the most important parameters for achieving high data transfer speed is the number of transmitter antennas. On the other hand, by increasing the number of transmitter antennas, the energy consumption of the system increases. By selecting the optimal number of antennas and the optimal power of users with the help of impact pilot reuse sequences, the EE of system is maximized [6]. The method of Lagrange dual function is used to get the optimal number of antennas and the optimal power of users. The proposed algorithm uses repeated searches to find the optimal number of antennas and power to avoid computational complexity. The simulation results prove the authors' claim and show that EE reaches its maximum value.

In [7], the authors focus on maximizing the secure EE. The system in question is a massive MIMO system in which an eavesdropper (EVA) destroys the confidential information of the system. To barricade this degradation, a cell division technique is proposed. In this technique, users are divided into two separate groups and appropriate transmission power is allocated to them according to location of users. A practical algorithm is also proposed that calculates the power of users in each group individually. The simulation results also show that the secure EE is optimized and the cell division technique improves the system performance. One of the strengths of the proposed algorithm is that in the presence of an EVA, it provide better results than the equal power allocation method.

Since the cell division method and the optimal selection of antennas improve performance of the system, we are motivated to determine which method is better to use in various situations by considering different precoding schemes for these two methods. The research work of this paper are summarized below.

\begin{itemize}
	\item
	
	This paper studies the operation of a massive MIMO system. An EVA also tries to corrupt system's information. Our purpose in this paper is to maximize secure EE. The maximization problem is under two constraints, which include the maximum transmission power and the minimum secure rate. The problem is solved by using linear programming and the Lagrange dual function.
	
	\item
	
	Two policies are used to achieve drawn goal, which include the cell division and optimal selection of the antennas. In the cell division technique, users are divided into two groups of far users and close users, which maximum transmission power is divided between these two groups based on their location. In the antenna selection mode, due to the large number of massive MIMO antennas, turning some of them off that are not cost-effective can improve the system performance. The MRT and ZF precoding schemes are employed for each method.
	
	\item
	Four iterative algorithms are designed and presented. In summary, Algorithm 1 does not use any of the methods, Algorithm 2 is related to the cell division method, Algorithm 3 is based on the strategy of selecting optimal antennas, and Algorithm 4 uses a hybrid method. The simulation results are given using numerical tests on three proposed algorithms, and different scenarios are compared with each other. Researchers can select the appropriate method for a similar system models based on their requirements.
	
\end{itemize}

The outline of this paper consists of 5 sections. Section II is dedicated to the introducing of the proposed system model, and formulates relations such as the received signal in each of the precoding schemes. In section III, we establish the optimization problem and solve it utilizing two methods, including the cell division technique and optimal antenna selection. Then, proposed algorithms are presented. Section IV analyzes obtained numerical results and compares different scenarios. Finally, section V discusses the conclusions of the paper and offers some future research topics.

Throughout the paper, parameters written by lowercase letters have scalar values, lowercase boldface letters represent vectors, and also uppercase boldface letters show matrices. Furthermore, $\Bbb{E}$$\left\lbrace .\right\rbrace $ is averaging operation, $\left( .\right) ^T$ shows transpose, $\left( .\right) ^H$ represents Hermitian, and $tr \left\lbrace .\right\rbrace $ is trace operation. Besides, $\Vert .\Vert $ denotes 2-norm of a vector, and $x^+ = \max \left( 0, x\right) $. Also, $\Re$ generates a matrix with standard Gaussian random variable elements, and $\lceil x \rceil$ shows the ceil of a real number $x$.

\section {System model}

The proposed system model is graphically depicted in Figure 1. The system consists of a BS equipped with $M$ antennas, and $K$ user with single antenna distributed around it. Furthermore, there is a single antenna EVA among users, which is not known to the BS and other users. In the cell division method, users are divided into two groups including $k_c$ central users and $k_e=k-k_c$ edge users. Also, $D$ is radius of the cell.

\begin{figure}
	\centering
	\includegraphics[scale=0.4]{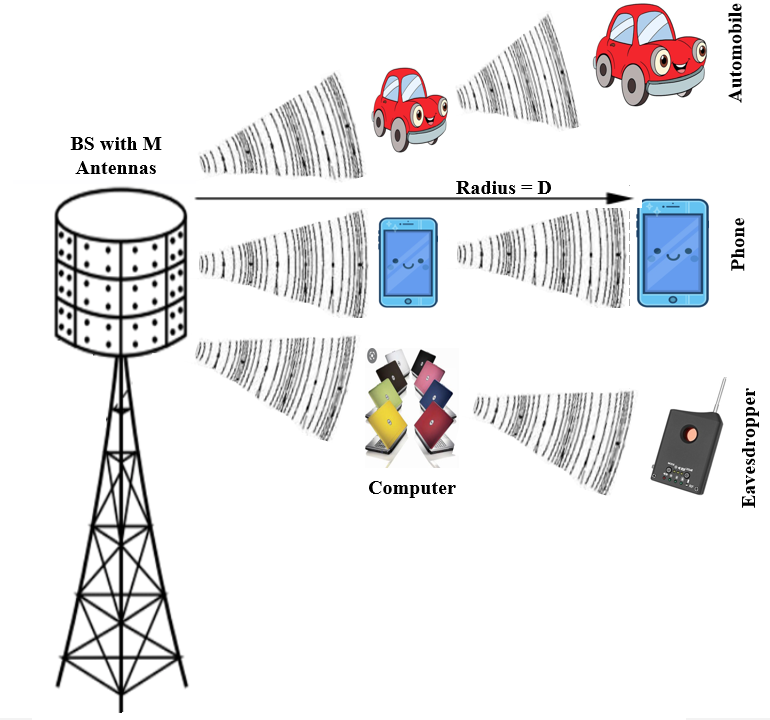}
	\caption{System model.} 
\end{figure}

\subsection {Channel model}

Different models have been considered for the analysis of telecommunication channels. The channel considered in this paper has the Rayleigh fading model. In this model, there exist many reflected paths between the transmitter and receiver with random gains, which are independent and identical random variables with the Gaussian distribution. These gains are real and imaginary parts of complex coefficients in a Rayleigh fading channel. The telecommunication channel between the BS and users is formulated as follows [8],

\begin{equation}
\textbf{C}=\textbf{H} \, \textbf{G}^{1/2},
\end{equation}
where,
\begin{equation}
{\textbf{G}}= \begin{bmatrix}

\zeta_1 & 0 & \dots & 0 \\

0 & 	\zeta_2 & \dots & 0 \\

\vdots & 	\vdots & 	\ddots & 	\vdots\\

0 & 0 & \dots & \zeta_K\\

\end{bmatrix}.
\end{equation}
In (2), $\zeta_k$ portends the path-loss and shadow fading effects for $k$-th user, simultaneously. Furthermore, we have,
\begin{equation}
\textbf{H}=\frac{1}{\sqrt 2} \left[  \Re (M,K)+\sqrt{-1}\hspace*{0.2cm} \Re (M,K)\right].
\end{equation}
Note that, $k \in \left\lbrace 1, 2 , \dots , K\right\rbrace $ and $m \in \left\lbrace 1, 2 , \dots, M\right\rbrace $.

Since the CSI is considered to be imperfect between the BS and users, an estimation of the ideal channel should be obtained. The telecommunication channel between the BS and users is formulated as follows [9],

\begin{equation}
\textbf{c}_k=\delta \hspace*{0.1cm} \hat{\textbf{c}}_k+\sqrt{1-\delta^2} \hspace*{0.15cm}\textbf{x}_k,
\end{equation}
where, $\textbf{c}_k$ represents the $k$-th column of the ideal channel matrix  $\textbf{C}$, and $\delta$ represents the accuracy of the channel estimation. Note that if $\delta = 1$, then the CSI is perfect. $\hat{\textbf{c}}_k$ is an estimation of the ideal channel vector with zero mean and variance of 1, i.e., its distribution is $\cal{CN}$$(0,1)$. The estimation error vector is also denoted by $\textbf{x}_k$, which has zero mean and a variance of 1. Since the channel estimation is calculated using the MMSE estimator, it is concluded that $ \hat{\textbf{c}}_k$ and $\textbf{x}_k$ are independent of each other, according to its orthogonality property.

We do not consider the power of each user equally allocated and we seek to calculate optimal power of each user. The signal received by the $k$-th user is equal to,

\begin{equation}
{y}_k=\sqrt{\upsilon p_k}\hspace*{0.15cm}\delta\hspace*{0.1cm} \hat{\textbf{c}}_k^H\textbf{b}_k s_k +\sqrt{\upsilon}\hspace*{0.1cm}\delta \sum_{j=1 , j \neq k}^{K}\sqrt{p_j}\hspace*{0.1cm} \hat{\textbf{c}}_k^H\textbf{b}_j s_j + \sqrt{\upsilon(1-\delta^2)}\sum_{i=1}^{K} \sqrt{p_i}\hspace*{0.1cm}\textbf{x}_k^H\textbf{b}_i s_i +n_k,
\end{equation}
where, the transmission power of the $k$-th user is represented by $p_k$, and $\textbf{b}_k$ represents $k$-th column of the precoding matrix $\textbf{B}$, which is different for the MRT and the ZF schemes. $\upsilon$ also represents the normalization coefficient, which is calculated by $\upsilon = \frac {1}{\Bbb{E}\left\lbrace tr\left( \textbf{B}^H \textbf{B}\right) \right\rbrace  }$. Furthermore, $s_k$ is symbol of the $k$-th user, which has two important properties, including $\Bbb{E}\left\lbrace  s_k \right\rbrace = 0$ and $\Bbb{E}\left[ \vert s_k \vert ^2\right]  = 1$. The additive white Gaussian noise (AWGN) is denoted by $n_k$, has zero mean and  variance of 1, i.e., $n_k \sim \cal{CN}$$(0,1)$.

According to (5), the signal-to-interference plus-noise ratio (SINR) of each user is formulated as follows,

\begin{equation}
\xi_k =  \frac {\upsilon p_k \delta^2 \vert\hat{\textbf{c}}_k^H \hat{\textbf{b}}_k\vert ^2 }{\upsilon \delta^2 \sum_{j=1 , j \neq k}^{K} p_j \vert\hat{\textbf{c}}_k^H \hat{\textbf{b}}_j\vert ^2 + \upsilon \left( 1-\delta^2\right)  \sum_{i=1}^{K} p_i \vert {\textbf{x}}_k^H \hat{\textbf{b}}_i\vert ^2+1}.
\end{equation}

\subsection {SINR for the MRT precoding}
By considering the MRT precoding scheme, it is concluded that $\textbf{b}_k = \hat{\textbf{c}}^H_k$. Thus, the SINR is formulated as follows,

\begin{equation}
\xi^{\text{MRT}}_k =  \frac {\upsilon p_k \delta^2 \vert\hat{\textbf{c}}_k^H \hat{\textbf{c}}_k\vert ^2 }{\upsilon\delta^2 \sum_{j=1 , j \neq k}^{K} p_j \vert\hat{\textbf{c}}_k^H \hat{\textbf{c}}_j\vert ^2 + \upsilon \left( 1-\delta^2\right)  \sum_{i=1}^{K} \vert {\textbf{x}}_k^H \hat{\textbf{c}}_i\vert ^2+1 }.
\end{equation}

Using [10], the SINR is equal to,

\begin{equation}
\xi^{\text{MRT}}_k = \frac {\delta p_k  M}{\left( K-\delta^2\right) p_k + K}.
\end{equation}

\subsection {SINR for the ZF precoding}

The $\textbf{B}$ matrix is utilized in the ZF precoding, and we have,

\begin{equation}
\textbf{B}=\hat{\textbf{C}}^H \left( \hat{\textbf{C}}\hspace*{0.15cm}\hat{\textbf{C}}^H\right)^{-1},
\end{equation}
When the used precoding is ZF, its properties, which include $\hat{\textbf{c}}_k \textbf{b}_k =1$ and $\hat{\textbf{c}}_k \textbf{b}_i =0$  for $1\le i \le K , i\neq k $, can be used and the SINR is formulated as follows,

\begin{equation}
\xi^{\text{ZF}}_k=\frac{\upsilon \delta^2 p_k}{ \sum_{j=1 ,}^{K} p_k \vert\textbf{x}_k\textbf{b}_i\vert ^2+1 } .
\end{equation}

Hence, the SINR is [10],

\begin{equation}
\xi^{\text{ZF}}_k= \frac{\upsilon \delta^2 p_k}{\upsilon \left( 1- \delta^2\right) \frac {1}{\left( M-1\right) \left( M-2\right)} p_k+1}.
\end{equation}

\subsection {SINR for the EVA}

As the worst case scenario is assumed for the system, the EVA can easily disrupt the system. Hence, the CSI of the EVA is considered to be perfect. On the other hand, EVA uses the MRT precoding and the received signal is as follows,

\begin{equation}
\textbf{y}^e_k = \sqrt {\zeta_e p_k} \textbf{C}_e \textbf{C}^H_e s_k + \sqrt {\zeta_e}\textbf{C}^H_e \sum_{i=1 ,i\neq k}^{K} p_i \textbf{C}_e s_i + n_e,
\end{equation} 
where, $n_e\sim \cal{CN}$$(0,1)$ is the AWGN at the EVA’s channel.

Using [7], the SINR is formulated as follows,

\begin{equation}
\xi^{e,k}= \frac {\zeta_e p_k}{\zeta_e\sum_{i=1 ,i\neq k}^{K} p_i +1}.
\end{equation}

\subsection {Secure data rate and secure EE}

A lower bound for the data rate can be written as follows [11]-[12],

\begin{equation}
R_k = \log_2 (1 + \xi_k )  > \tilde{R}_k =   \log_2  \xi_k , 
\end{equation}
and hence,
\begin{equation}
\tilde{R}_{e,k} =  \log_2 \left(\frac {\zeta_e p_k}{\zeta_e\sum_{i=1 ,i\neq k}^{K} p_i +1}\right) .
\end{equation}
As a result, we proceed as follows to obtain the secure data rate definition,

\begin{equation}
\tilde{R}^{\text{sec}}_k \approx \left[ \tilde{R}_k - \tilde{R}_{e,k}\right] ^+.
\end{equation}
This approximation is reasonable in the case of using a massive number of antennas at the BS because of the channel hardening property.

Considering that secure EE is equal to the ratio of secure SE to the transmission and consumed power, with increasing power, the secure EE decreases, so there is always a compromise between the secure EE and total power. The definition of secure EE is as follows [13],

\begin{equation}
q_{\text{sec}}=\frac{ \sum_{k=1}^{K} \tilde{R}^{\text{sec}}_k}{ \sum_{k=1}^{K} p_k + M P_c},
\end{equation}
where, the $P_c$ is the constant power consumption by the circuit.

\section {Proposed schemes and proposed algorithms}

First, the optimization problem should be formulated mathematically and the related constraints must be stated. The constraints on the problem of secure EE maximization include maximum transmission power $P_{\max}$, and minimum secure rate. The maximization problem is established as follows,

\begin{subequations}
	\begin{equation}
	\underset{\left\lbrace p_1,p_2,\dots,p_k\right\rbrace }{\max} \quad q_{\text{sec}} = \frac{\sum_{k=1}^{K} \tilde{R}^{\text{sec}}_k}{ \sum_{k=1}^{K} p_k + M P_c}
	\end{equation}
	\begin{equation}
	C1 \quad : \quad \sum_{k=1}^{K} p_k \le P_{\max},
	\end{equation}
	\begin{equation}
	C2 \quad : \quad \tilde{R}^{\text{sec}}_k \ge R_{k,\min},
	\end{equation}
\end{subequations}
where, $R_{k,\min}$ indicates the minimum secure rate of $k$-th user needs to meet a QoS.

The optimization problem in (18) has two main issues that make it difficult to solve. They are the non-convexity of the objective function and the issue due to having two constraints. In order to resolve them, nonlinear programming features and the Lagrange dual function are used.

Using the nonlinear programming features and the lower bound secure data rate of $ \tilde{R}^{\text{sec}}_k$, the optimization problem can be rewritten convexly. With these assumptions, the optimization problem in (18) is equivalent to [14],

\begin{subequations}
	\begin{equation}
	\underset{\left\lbrace p_1,p_2,\dots,p_k\right\rbrace}{\max} \left( \sum_{k=1}^{K} \tilde{R}^{\text{sec}}_k -q^* \left( \sum_{k=1}^{K} p_k +M P_{c}\right) \right)  
	\end{equation}
	\begin{equation}
	C1 \quad : \quad \sum_{k=1}^{K} p_k \le P_{\max},
	\end{equation}
	\begin{equation}
	C2 \quad : \quad \tilde{R}^{\text{sec}}_k \ge R_{k,\min},
	\end{equation}
\end{subequations}
where, $q^*$ represents maximum secure EE value.

\subsection {The proposed schemes}

In this section, the optimization problem is solved without using the method of cell division and antenna selection, and the optimal power of users are calculated. Since the problem of optimization is constrained, it is turned into an unconstrained problem. We have [15],

\begin{align}
\phi (\textbf{P}, \Psi, \boldsymbol{\Gamma}) = &- \left [\sum_{k=1}^{K} \tilde{R}^{\text{sec}}_k -q^* \left( \sum_{k=1}^{K} p_k +  M P_{c}\right)  \right]  \\ \nonumber&-  \Psi \left (P_{\max} - \sum_{k=1}^{K} p_k \right)
-\sum_{k=1}^{K} \Gamma_k \left( \tilde{R}^{\text{sec}}_k -R_{k,\min}\right) .
\end{align}
By doing some mathematical simplifications,

\begin{align}
\phi (\textbf{P},\Psi, \boldsymbol{\Gamma}) = &- \left [\sum_{k=1}^{K} \left( 1+ \Gamma_k\right) \tilde{R}^{\text{sec}}_k -q^* \left( \sum_{k=1}^{K} p_k  + M P_{c}\right)  \right]  \\ \nonumber&- \Psi \left (P_{\max} - \sum_{k=1}^{K} p_k \right) +\sum_{k=1}^{K}  \Gamma_k R_{k,\min},
\end{align}
where, $\textbf{P}$ is acceptable power set of users. Also $\Psi > 0$ and $\Gamma_k > 0$ are the coefficient of Lagrange function corresponding to the maximum transmission power and the minimum secure rate for users, respectively.

\subsubsection {The proposed scheme for the MRT}

Here, we consider the MRT precoding and its related rate that has been obtained previously. To obtain the optimal power, we derive the Lagrange dual function with respect to the power of users and set it to be zero. The result is,

\begin{equation}
\frac{\partial \phi}{\partial p_k}=0.
\end{equation}
After some mathematical simplifications, we come to the following relation,

\begin{equation}
\frac{\left( 1+\Gamma_k\right)  \left( K-\delta^2\right)}{\left(\left( K-\delta^2\right) p_k + K\right) \ln 2} - \sum_{j=1 ,j\neq k}^{K} \frac { 1+\Gamma_j}{\left(\sum_{i=1 ,i\neq j}^{K} p_i + \frac {1}{\zeta_e} \right) \ln 2 } + q + \Psi = 0.
\end{equation}
Therefore, the optimal power of users is obtained as follows,

\begin{equation}
p_k = \frac{\left( 1+\Gamma_k\right)}{ \left( \sum_{j=1 ,j\neq k}^{K} \frac {1+\Gamma_j}{\left(\sum_{i=1 ,i\neq j}^{K} p_i + \frac {1}{\zeta_e} \right) \ln 2 } - q - \Psi\right)\ln 2 } - \frac {K}{ K-\delta^2}. 
\end{equation}

\subsubsection {The proposed scheme for the ZF}

If the ZF precoding scheme is employed, the rate obtained for the ZF is used. To obtain the optimal power, we similarly derive the Lagrange dual function and set it to be zero. Thus,

\begin{equation}
\frac{\upsilon \left( 1-\delta^2\right) \frac {1}{\left( M-1\right) \left( M-2\right)} \left( 1+\Gamma_k\right) }{\left(\upsilon \left( 1-\delta^2\right) \frac {1}{\left( M-1\right) \left( M-2\right)} p_k + 1\right) \ln 2 }- \sum_{j=1, j\neq k}^{K} \frac {1+\Gamma_j}{\left( \sum_{i=1, i\neq j}^{K} p_i +  \frac{1}{\zeta_e}\right)\ln 2 } + q +\Psi=0,
\end{equation}
and then,
\begin{equation}
p_k=\frac {1+\Gamma_k}{\left( \sum_{j=1, j\neq k}^{K} \frac {1+\Gamma_j}{\left( \sum_{i=1, i\neq j}^{K} p_i +  \frac{1}{\zeta_e}\right)\ln 2 }-q-\Psi\right)\ln 2 }-\frac{1}{\upsilon \left( 1-\delta^2\right) \frac {1}{\left( M-1\right) \left( M-2\right)}}.
\end{equation}

Algorithm 1 provides a numerical solution for the proposed scheme.In this algorithm, $n$ represents the number of iterations of the algorithm. $\iota$ and $ \varpi$ are also positive steps used to update Lagrange coefficients [16].

\begin{table}[h!]
	
	\begin{tabular}{l}
		\hline
		\\
		\textbf{Algorithm 1}: Find the optimal power with the proposed schemes.   \\
		\\ \hline
		1. Initialize the Lagrangian multipliers $(\Psi ^{(0)}$, $\Gamma ^{(0)}_k$), Feasible power of users,\\ Minimum rate of users;\\
		2.  Calculate the initial EE $q^{(0)}_{\text{sec}}=\frac{\sum_{k=1}^{K} \tilde{R}^{\text{sec}^{(0)}}_k}{ \sum_{k=1}^{K} p^{(0)}_k + M P_c}$\\
		3. While $\left( \sum_{k=1}^{k} \tilde{R}_k^{(n)}-q^{(n)}_{\text{sec}}(\sum_{k=1}^{k}p^{(n)}_k+MP_c)>\beta\right) $ do\\
		4.$\,\,\,\,\,\,$ {for} $i=1: K$ {do}\\
		$\quad\quad\,\,\,\,\,\,\,\,\,\,\,\,\,$ MRT: Calculate $p_k^{(n+1)}$ according to formula (24),\\
		$\quad\quad\,\,\,\,\,\,\,\,\,\,\,\,\,\,\,\,$ZF: Calculate $p_k^{(n+1)}$ according to formula (26),\\
	    $\,\,\,\,\,\,\,\,\,\,\,\,$ {end for}\\ 
	    5. Update\\
	    $p^{(n+1)}_k=p^{(n)}_k$\\
		$\Psi^{(n+1)}= \max(0,\Psi_1^{(n)}-\iota(P_{\max}-\sum_{k=1}^{K} p_k^{(n)}))$\\
	     $\Gamma^{(n+1)}_k= \max(0,\Gamma^{(n)}_k-\varpi(\tilde{R}_{\text{sec},k}^{(n)}-R_{k,\min}))$\\
	    $q^{(n+1)}_{\text{sec}}={\frac{\sum_{k=1}^{K} \tilde{R}^{\text{sec}}_k}{ \sum_{k=1}^{K} p^{(n)}_k + M P_c}}$\\
	    $n=n+1$\\
	    end while\\
	    End\\ \hline 
	\end{tabular}
\end{table}

\subsection {Cell division schemes}

One of the proposed methods to improve the system performance is the cell division strategy. In this method, users are divided into two groups: central users ($k_c$) and edge users ($k_e$). Depending on the location of users in the group, the maximum transmission power is divided between them. For this reason, the maximum transmission power constraint turns into two constraints which are defined for each group of users. The optimization problem in this method is established as follows,

\begin{subequations}
	\begin{equation}
	\underset{\left\lbrace p_1,p_2,\dots,p_k\right\rbrace}{\max} \left( \sum_{k=1}^{K} \tilde{R}^{\text{sec}}_k -q \left( \sum_{k\in \text{First group}} p_k + \sum_{k\in \text{Second group}} p_k +M P_c\right) \right)  
	\end{equation}
	\begin{equation}
	C1  : \quad \sum_{k\in \text{First group}} p_k  = \Lambda P_{\max},
	\end{equation}
	\begin{equation}
	C2  :  \sum_{k\in \text{Second group}} p_k \le \left( 1-\Lambda \right) P_{\max},
	\end{equation}
	\begin{equation}
	C3 :   {\tilde R}^{\text{sec}}_k \ge R_{k,\min},
	\end{equation}
\end{subequations}
where, $\Lambda$ is the coefficient of maximum transmission power in the first group, and it is calculated as $\Lambda = \frac {\sum_{k\in \text{First group}} d_k}{\sum_{k=1}^{K} d_k}$.

The Lagrange dual function is used to eliminate the problem constraints. Thus, we have,

\begin{align}
\phi (\textbf{P},\Psi_1,\Psi_2, \boldsymbol{\Gamma}) = &- \left [\sum_{k=1}^{K} \tilde{r}^{\text{sec}}_k -q \left( \sum_{k\in \text{First group}} p_k + \sum_{k\in \text{Second group}} p_k + M P_{c}\right)  \right]  \\ \nonumber&- \Psi_1 \left (\Lambda P_{\max} - \sum_{k\in \text{First group}} p_k \right) \\ \nonumber& - \Psi_2 \left (\left( 1-\Lambda\right) P_{\max}-\sum_{k\in \text{Second group}} p_k  \right)&
\\ \nonumber&-\sum_{k=1}^{K} \Gamma_k \left( \tilde{R}^{\text{sec}}_k -R_{k,\min}\right) ,
\end{align}
where, $\Psi_1$ and $\Psi_2$ are respectively coefficients of the Lagrange function corresponding to the maximum transmission power constraints in the first and second groups.

\subsubsection {Cell division scheme for the MRT}

After calculating the derivative and putting it to be zero, the optimal power of users in the first area is obtained,
\begin{equation}
\frac{\left( 1+\Gamma_k\right)  \left( K-\upsilon^2\right)}{\left(\left( K-\upsilon^2\right) p_k + K\right) \ln 2} - \sum_{j=1 ,j\neq k}^{K} \frac {1+\Gamma_j}{\left(\sum_{i=1 ,i\neq j}^{K} p_i + \frac {1}{\zeta_e} \right) \ln 2 } + q + \Psi_1 = 0,
\end{equation}
and hence,

\begin{equation}
p_k = \frac{\left( 1+\Gamma_k\right)}{ \left( \sum_{j=1 ,j\neq k}^{K} \frac {1+\Gamma_j}{\left(\sum_{i=1 ,i\neq j}^{K} p_i + \frac {1}{\zeta_e} \right) \ln 2 } - q - \Psi_1\right)\ln 2 } - \frac {K}{ K-\upsilon^2}. 
\end{equation}
Similarly, for the second area we get,

\begin{equation}
p_k = \frac{\left( 1+\Gamma_k\right)}{ \left( \sum_{j=1 ,j\neq k}^{K} \frac {1+\Gamma_j}{\left(\sum_{i=1 ,i\neq j}^{K} p_i + \frac {1}{\zeta_e} \right) \ln 2 } - q - \Psi_2\right)\ln 2 } - \frac {K}{ K-\upsilon^2}. 
\end{equation}

\subsubsection {Cell division scheme for the ZF}

By taking similar steps, for the first area, it is obtained that,
\begin{equation}
\frac{\upsilon \left( 1-\delta^2\right) \frac {1}{\left( M-1\right) \left( M-2\right)} \left( 1+\Gamma_k\right) }{\left(\upsilon \left( 1-\delta^2\right) \frac {1}{\left( M-1\right) \left( M-2\right)} p_k + 1\right) \ln 2 }- \sum_{j=1, j\neq k}^{K} \frac {1+\Gamma_j}{\left( \sum_{i=1, i\neq j}^{K} p_i +  \frac{1}{\zeta_e}\right)\ln 2 } + q +\Psi_1=0.
\end{equation}
Then, the optimal power of users in the first area is equal to,

\begin{equation}
p_k=\frac {1+\Gamma_k}{\left( \sum_{j=1, j\neq k}^{K} \frac {1+\Gamma_j}{\left( \sum_{i=1, i\neq j}^{K} p_i +  \frac{1}{\zeta_e}\right)\ln 2 }-q-\Psi_1\right)\ln 2 }-\frac{1}{\upsilon \left( 1-\delta^2\right) \frac {1}{\left( M-1\right) \left( M-2\right)}}.
\end{equation}
Equivalently, we can write for the second area,

\begin{equation}
p_k=\frac {1+\Gamma_k}{\left( \sum_{j=1, j\neq k}^{K} \frac {1+\Gamma_j}{\left( \sum_{i=1, i\neq j}^{K} p_i +  \frac{1}{\zeta_e}\right)\ln 2 }-q-\Psi_2\right)\ln 2 }-\frac{1}{\upsilon \left( 1-\delta^2\right) \frac {1}{\left( M-1\right) \left( M-2\right)}}.
\end{equation}

Finally, we propose the Algorithm 2 as a numerical solution for the cell division scheme and its obtained results are reported in the next section.

\begin{table}[h!]
	
	\begin{tabular}{l}
		\hline
		\\
		\textbf{Algorithm 2}: Find the optimal power with the cell division technique.   \\
		\\ \hline
		1. Initialize the Lagrangian multipliers $(\Psi_1 ^{(0)}$,$\Psi_2 ^{(0)}$, $\Gamma ^{(0)}_k$), Feasible power of users,\\ Minimum rate of users; \\
		2.  Calculate the initial secure EE $q^{(0)}_{\text{sec}}=\frac{\sum_{k=1}^{K} \tilde {R}^{\text{sec}(0)}_{k}}{\sum_{k \in \text{First group}} p^{(0)}_k + \sum_{k \in \text{Second group}} p^{(0)}_k + M P_c}$\\
		3. While $\left( \sum_{i=1}^{K} \tilde {R}_i^{\text{sec}(n)}-q^{(n)}\left( \sum_{k \in \text{First group}} p^{(n)}_k + \sum_{k \in \text{Second group}} p^{(n)}_k + M P_c\right) >\beta\right) $ do\\
		4.$\,\,\,\,\,\,$ {for} $k=1: K$ {do}\\
		$\,\,\,\,\,\,\,\,\,\,\,\,\,\,$ {if} $k \in \text{First group}$ {do}\\
		$\quad\quad\,\,\,\,\,\,\,\,\,\,\,$MRT: Calculate the $p_k^{(n+1)}$ according to formula (30)\\
		$\quad\quad\,\,\,\,\,\,\,\,\,\,\,\,\,\,\,\,$ZF: Calculate the $p_k^{(n+1)}$ according to formula (33)\\
		$\,\,\,\,\,\,\,\,\,\,\,\,\,\,$ else\\
		$\quad\quad\,\,\,\,\,\,\,\,\,\,\,\,\,$MRT: Calculate the $p_k^{(n+1)}$ according to formula (31)\\
		$\quad\quad\,\,\,\,\,\,\,\,\,\,\,\,\,$ZF: Calculate the $p_k^{(n+1)}$ according to formula (34)\\
		$\,\,\,\,\,\,\,\,\,\,\,\,\,\,$ end if\\
		$\,\,\,\,\,\,\,\,\,\,\,\,$ {end for}\\ 
		5. Update\\
		$\Psi_1^{(n+1)}= \max(0,\Psi_1^{(n)}-\iota(P_{\max}-\sum_{k \in \text{First group}} p_k^{(n)}))$\\
		 $\Psi_2^{(n+1)}= \max(0,\Psi_2^{(n)}-\iota(P_{\max}-\sum_{k \in \text{Second group}} p_k^{(n)}))$\\
		$\Gamma^{(n+1)}_k= \max(0,\Gamma^{(n)}_k-\varpi(\tilde{R}_{\text{sec},k}^{(n)}-R_{k,\min}))$\\
		$q^{(n+1)}_{\text{sec}}=\frac{\sum_{k=1}^{k} \tilde {R}^{\text{sec}(n)}_{k}}{\sum_{k \in \text{First group}} p^{(n)}_k + \sum_{k \in \text{Second group}} p^{(n)}_k + M P_c}$\\
		$n=n+1$\\
		end while\\
		End\\ \hline 
	\end{tabular}
\end{table}

\subsection {Antenna selection schemes}

Since the number of antennas in massive MIMO systems is large, it is very important to select antennas that are not economical to utilize. If the number of antennas decreases, then the constant power of the circuit decreases and the secure EE increases. In our proposed scheme, the number of optimal antennas is calculated using the Hessian matrix and the Newton method.

In order to calculate the number of optimal antennas, we first build the optimization problem as follows [17],

\begin{equation}
\underset{M}{\max}  \quad \frac{ B\,\sum_{k=1}^{K} \tilde{R}^{\text{sec}}_k}{ \sum_{k=1}^{K} p_k + M P_{c}},
\end{equation}
where, $B$ is the bandwidth of the system. This optimization problem is rewritten as follows according to the set of acceptable antennas ($\nu$),

\begin{equation}
\underset{M \in \nu}{\max} \quad \bar{\Theta}\left( M\right)  = \frac{\Omega_1\left( M\right) }{\Omega_2\left( M\right) },
\end{equation}
where, $\Omega_1=\sum_{k=1}^{K} \tilde{R}^{\text{sec}}_k $ is a concave function and $\Omega_2 = \sum_{k=1}^{K} p_k + M P_{c} $ is a convex one. We can consider a constraint to this optimization problem, i.e.,

\begin{align}
&\underset{M \in \nu }{\max} \quad \bar{\Theta}\\ \nonumber
&\text{subject to} \quad :  \frac{\Omega_1\left( M\right) }{\Omega_2\left( M\right) } - \bar{\Theta}  \ge 0,
\end{align}
where, $\bar{\Theta}$ is the optimal value of the secure EE. We consider a set $\varepsilon$ such that $\nu \subseteq \varepsilon$, where $\varepsilon$ shows antenna set of the BS. 

Using the Newton's method, the optimization problem is reformulated as follows [18],

\begin{equation}
\mu \left(\bar{\Theta}\right) = \underset{M}{\max} \quad \Omega_1\left( M\right) - \bar{\Theta} \Omega_2\left( M\right) .
\end{equation}
Hence $\Omega_1\left( M\right)$ is a concave function and $\bar{\Theta}\, \Omega_2\left( M\right)$ is an affine function, thus $\Omega_1\left( M\right) - \bar{\Theta} \Omega_2\left( M\right)$ is a concave function. The Newton's method is used to obtain the optimal secure EE value. Thus we have,

\begin{equation}
\bar{\Theta}_{n+1} =  \bar{\Theta}_{n} - \frac {\mu\left(\bar{\Theta}_{n}\right)}{\mu^{'}\left(\bar{\Theta}_{n}\right)} = \frac {\Omega_1\left( M^{opt}\right)}{\Omega_2\left( M^{opt}\right)}.
\end{equation}

First, the initial value of the secure EE must be calculated, the value of which is called $\bar{\Theta}_0$. This initial value should satisfy $\mu\left(\bar{\Theta}_0\right) \le 0 $. By having the optimal number of antennas $(M^{opt})$, $\mu\left(\bar{\Theta}_n\right)  $ can be obtained.


Utilizing the secure rate of the MRT precoding, the optimal number of antennas is obtained as follows,

\begin{equation}
M^{opt} = \lceil \frac{B}{\bar{\Theta}_n P_c \ln 2}  \rceil,
\end{equation}
and using the secure rate of the ZF precoding, the optimal number of antennas is obtained as follows,

\begin{equation}
M^{opt} = \lceil \frac{B}{\bar{\Theta}_n P_c \ln 2}+K  \rceil.
\end{equation}
The Algorithm 3 is designed for solving the problem by antenna selection scheme, numerically.

\begin{table}[h!]
	
	\begin{tabular}{l}
		\hline
		\\
		\textbf{Algorithm 3}: Find the optimal power with antenna selection scheme.   \\
		\\ \hline
		1: Initialize the Lagrangian multipliers $(\Psi ^{(0)}$, $\Gamma ^{(0)}_k$), Feasible power of users, \\Minimum rate of users;\\
		
		2: Calculate the initial secure EE, $\bar{\Theta}^{(0)}=\frac{\sum_{k=1}^{K} \hat R^{\text{sec}(0)}_{k}}{\sum_{k=1}^{K}p^{(0)}_k+MP_c}$\\
		
		3: While $\left [\sum_{k=1}^{K}  \tilde{R}^{(n)}_{\text{sec},k} -{\bar{\Theta}}^{(n)} \left( \sum_{k=1}^{K} p_k^{(n)} +  M P_{c}\right)  \right] \ge \varrho $ do\\
		
		4: Do the antenna selection.\\
		$\,\,\,\,\,\,\,\,\,\,\,\,\,\,$ MRT: Use (40) to calculate the optimal number of antennas.\\
		$\,\,\,\,\,\,\,\,\,\,\,\,\,\,$ ZF: Use (41) to calculate the optimal number of antennas.\\
		
		5: for $k=1:K$ do\\
		$\,\,\,\,\,\,\,\,\,\,\,\,\,\,$ MRT: Calculate the $p_k^{(n+1)}$ according to formula (24).\\
		$\,\,\,\,\,\,\,\,\,\,\,\,\,\,\,\,$ZF: Calculate the $p_k^{(n+1)}$ according to formula (26).\\
		$\,\,\,\,\,\,$end for\\
		
		Update\\
		
		$p^{(n+1)}_k=p^{(n)}_k$.\\
		
		$\Psi^{(n+1)}=\max(0,\Psi^{(n)}-\iota(P_{\max}-\sum_{k=1}^{K} p_k^{(n)}))$.\\
		
		$\Gamma^{(n+1)}_k= \max(0,\Gamma^{(n)}_k-\tau(\tilde{R}_{\text{sec},k}^{(n)}-R_{k,\min}))$.\\
		
		$\bar{\Theta}_{n+1} =  \bar{\Theta}_{n} - \frac {\mu\left({\bar{\Theta}}_{n}\right)}{\mu^{'}\left({\bar{\Theta}}_{n}\right)}$.\\
		
		 $n=n+1$.\\
		 
		 end while\\
		 
		 End.\\\hline 
	\end{tabular}
\end{table}

\subsection {Joint antenna selection and cell division scheme}

In this subsection, we propose a combined method using the relations obtained in the previous subsections. Thereby, an iterative algorithm is designed for the combined method of optimal antenna selection and cell division technique. Algorithm 4 provides a numerical solution of the joint antenna selection and cell division scheme.

\begin{table}[h!]
	
	\begin{tabular}{l}
		\hline
		\\
		\textbf{Algorithm 4}: Find the optimal power with antenna selection and cell division scheme.   \\
		\\ \hline
		1: Initialize the Lagrangian multipliers $(\Psi ^{(0)}$, $\Gamma ^{(0)}_k$), Feasible power of users, \\Minimum rate of users;\\
		
		2: Calculate the initial secure EE, $\bar{\Theta}^{(0)}=\frac{\sum_{k=1}^{K} \hat R^{\text{sec}(0)}_{k}}{\sum_{k=1}^{K}p^{(0)}_k+MP_c}$\\
		
		3: While $\left [\sum_{k=1}^{K}  \tilde{R}^{(n)}_{\text{sec},k} -{\bar{\Theta}}^{(n)} \left( \sum_{k=1}^{K} p_k^{(n)} +  M P_{c}\right)  \right] \ge \varrho $ do\\
		
		4: Do the antenna selection.\\
		$\,\,\,\,\,\,\,\,\,\,\,\,\,\,$ MRT: Use (40) to calculate the optimal number of antennas.\\
		$\,\,\,\,\,\,\,\,\,\,\,\,\,\,$ ZF: Use (41) to calculate the optimal number of antennas.\\
		
		5: for $k=1:K$ do\\
		
		$\,\,\,\,\,\,\,\,\,\,\,\,\,\,$  if $k \in \text{First group}$ do\\
		
		$\,\,\,\,\,\,\,\,\,\,\,\,\,\,$ MRT: Calculate the $p_k^{(n+1)}$ according to formula (30).\\
		$\,\,\,\,\,\,\,\,\,\,\,\,\,\,\,\,$ZF: Calculate the $p_k^{(n+1)}$ according to formula (33).\\
		
		$\,\,\,\,\,\,\,\,\,\,\,\,\,\,$ else \\
		
		$\,\,\,\,\,\,\,\,\,\,\,\,\,\,$ MRT: Calculate the $p_k^{(n+1)}$ according to formula (31).\\
		$\,\,\,\,\,\,\,\,\,\,\,\,\,\,\,\,$ZF: Calculate the $p_k^{(n+1)}$ according to formula (34).\\
		
		$\,\,\,\,\,\,\,\,\,\,\,\,\,\,$ end if\\
		$\,\,\,\,\,\,$end for\\
		
		Update\\
		
		$p^{(n+1)}_k=p^{(n)}_k$.\\
		
		$\Psi_1^{(n+1)}= \max(0,\Psi_1^{(n)}-\iota(P_{\max}-\sum_{k \in \text{First group}} p_k^{(n)}))$\\
		$\Psi_2^{(n+1)}= \max(0,\Psi_2^{(n)}-\iota(P_{\max}-\sum_{k \in \text{Second group}} p_k^{(n)}))$\\
		$\Gamma^{(n+1)}_k= \max(0,\Gamma^{(n)}_k-\varpi(\tilde{R}_{\text{sec},k}^{(n)}-R_{k,\min}))$\\
		
		$\bar{\Theta}_{n+1} =  \bar{\Theta}_{n} - \frac {\mu\left({\bar{\Theta}}_{n}\right)}{\mu^{'}\left({\bar{\Theta}}_{n}\right)}$.\\
		
		$n=n+1$.\\
		
		end while\\
		
		End.\\\hline 
	\end{tabular}
\end{table}

\section {Simulation results and discussion}

In this section, the numerical results obtained from four proposed algorithms are analyzed. Numerical results related to the method of equal power allocation are also presented. The obtained results are based on implementation of algorithms in the MATLAB environment. In order to provide reliable and stable results, each algorithm is repeated 1000 times and average of their output is reported.

Some important parameters should be initialized in the simulation process, which their values are inspired by [19]-[21] and are given in Table I. The location of users and EVA is considered to be random.

 \begin{table}[!h]
	\centering
	\caption{Parameters which are used in simulations.}
	\begin{tabular}{|c|c|r|}
		\hline
		\textbf{Parameter} & \textbf{Value} \\
		\hline  
		Bandwdith & $120$ kHz \\
		\hline 
		Variance of log-normal shadow fading & $10$ dBm \\
		\hline 
		
		Noise spectral density & $-174$ dBm/Hz \\
		\hline  
		Constant power per antenna & $0.1$ Watt \\
		\hline 
		Positive step size of  $\iota$  & $0.01$ \\
		\hline 
		Positive step size of  $\varpi$ & $0.01$ \\
		\hline        
	\end{tabular} 
\end{table}

Figure 2 shows the numerical results obtained from Algorithm 1 and Algorithm 2. In this figure, how to change secure EE based on increasing the number of BS antennas is investigated. As shown in the figure, the cell division method improves system performance for both scenarios. Also, it has been shown that the system performance with ZF precoding is better than the MRT precoding. Both methods have better performance than the equal power allocation strategy.

\begin{figure}
	\centering
	\includegraphics[scale=0.6]{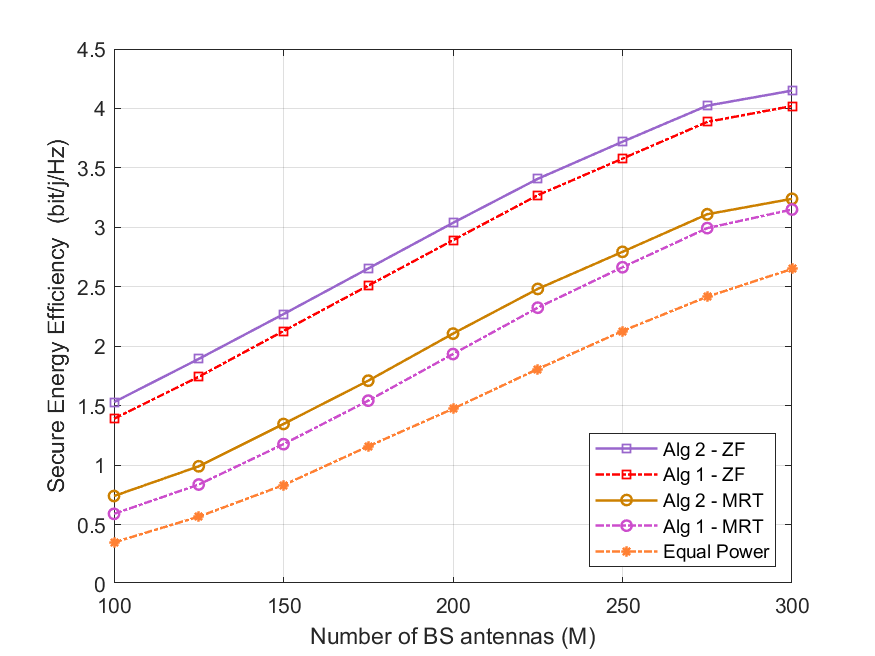}
	\caption{Secure energy efficiency versus the number of transmit antennas.} 
\end{figure}

Figure 3 shows that how the secure EE increases with increasing of the maximum transmission power for Algorithm 1 and Algorithm 2. Furthermore, the amount of improvement that the cell division method creates in the performance of the system is clearly shown. With the increase of maximum transmission power, the secure EE reaches a saturated state, and from one point onwards, with the increase of maximum transmission power, secure EE does not improve. In this analysis, the ZF precoding also performs better than the MRT precoding.

\begin{figure}
	\centering
	\includegraphics[scale=0.6]{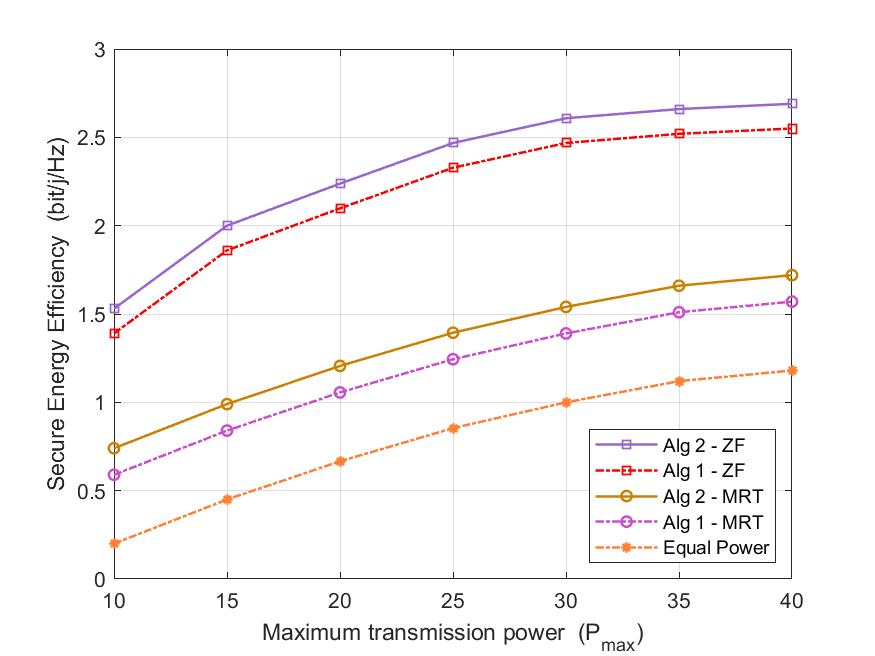}
	\caption{Secure energy efficiency versus the maximum transmission power ($P_{\max}$).} 
\end{figure}

Results for Algorithm 1 and Algorithm 3 are depicted in Figure 4. This figure shows the effect of selecting the optimal number of antennas on the system performance. It is concluded that with a smaller number of BS antennas, the optimal antenna selection scheme has no significant effect, while by increasing the number of BS antennas, the effect of this method is clearly observed and the difference in secure EE is much better compared to Algorithm 1.

\begin{figure}
	\centering
	\includegraphics[scale=0.6]{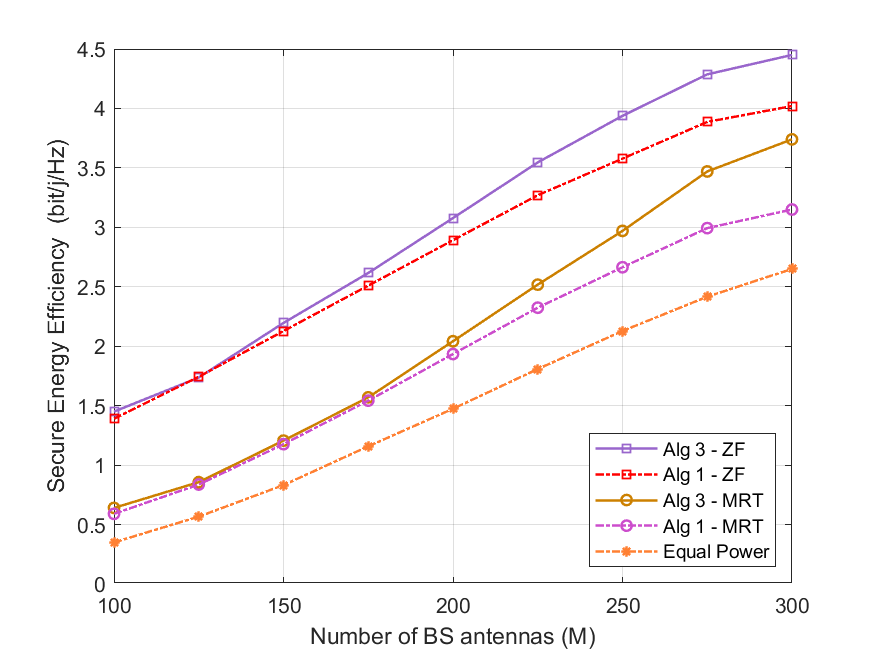}
	\caption{Secure energy efficiency versus the number of transmit antennas.} 
\end{figure}

In Figure 5, the numerical results obtained from Algorithm 1 and Algorithm 3 are analyzed. By increasing the maximum transmission power, the secure EE changes are examined. At low maximum transmission power values, there is not much improvement compared to the Algorithm 1, but with increasing maximum power, we see a significant improvement in the system performance. In this form, both proposed methods have better results than the equal power allocation scheme.

\begin{figure}
	\centering
	\includegraphics[scale=0.6]{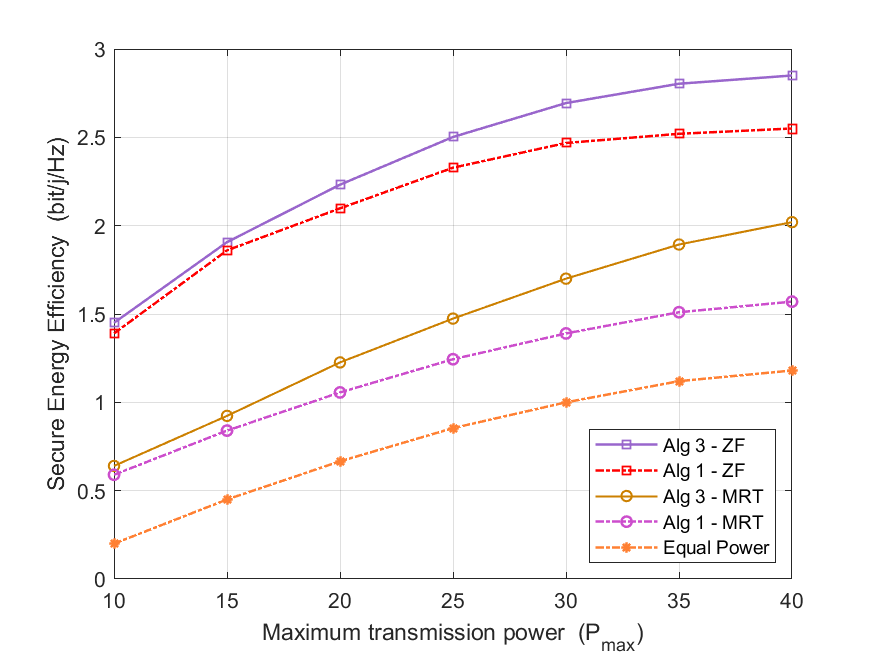}
	\caption{Secure energy efficiency versus the maximum transmission power ($P_{\max}$).} 
\end{figure}

Figure 6 compares two methods of the cell division and the optimal selection of antennas from the secure EE point of view. It is observed that at low values of maximum transmission power, the cell division method performs better, while for the higher maximum transmission power values, antenna selection method performs better. As a result, depending on the maximum transmission power, we can choose a proper method.

\begin{figure}
	\centering
	\includegraphics[scale=0.6]{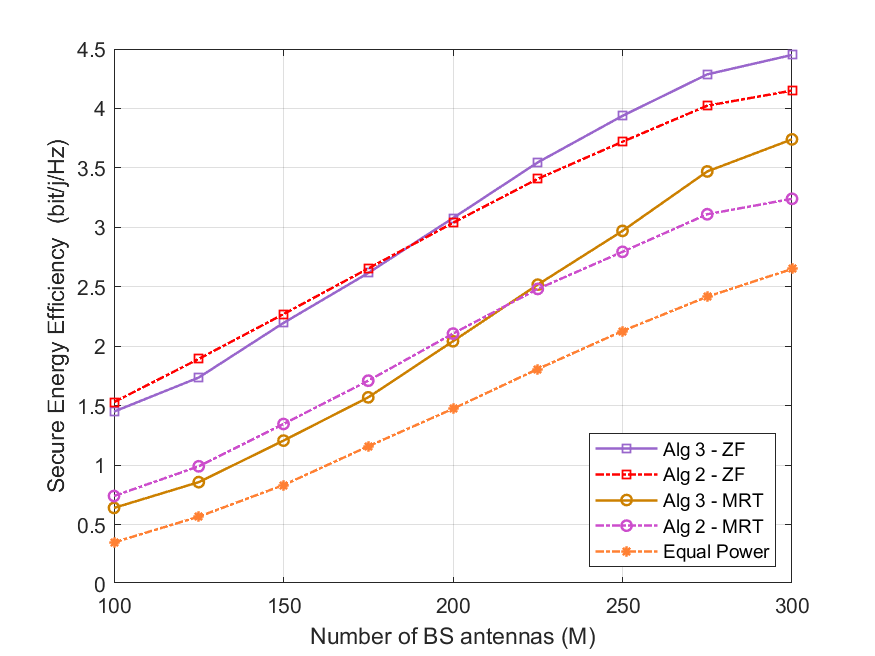}
	\caption{Secure energy efficiency versus the number of transmit antennas.} 
\end{figure}

The performance of secure EE is investigated by two methods of cell division and optimal selection of antennas by increasing the number of BS antennas in Figure 7. In this analysis, for the number of antennas less than 220 in the ZF precoding and for the number of antennas less than 225 in the MRT precoding, the cell division method has a better performance. Clearly, for the number of antennas more than the mentioned values, the optimal antennas selection scheme outperforms. Based on this analysis, researchers can easily decide which method provides the best performance for a specific system model, given the number of BS antennas.

\begin{figure}
	\centering
	\includegraphics[scale=0.6]{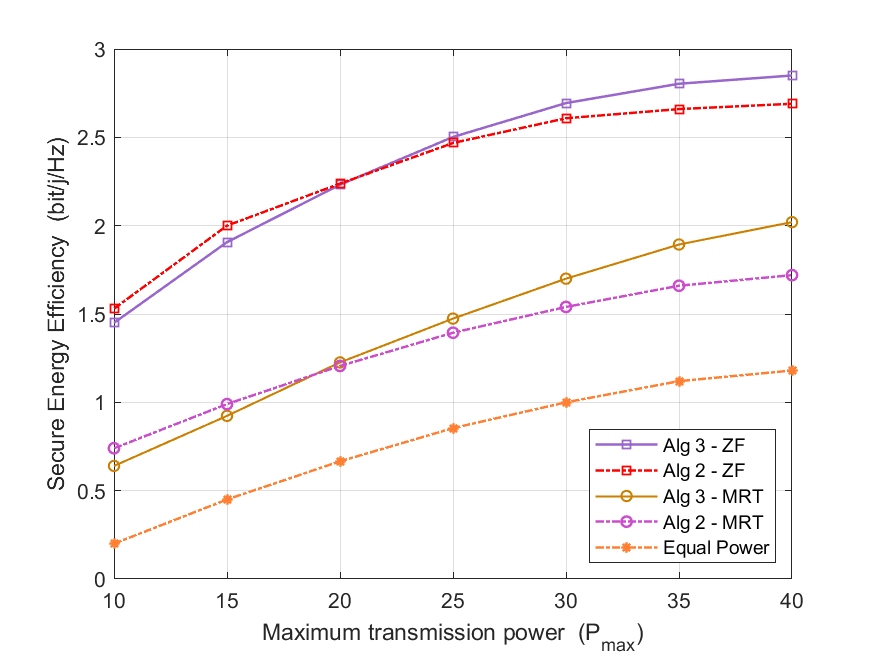}
	\caption{Secure energy efficiency versus the maximum transmission power ($P_{\max}$).} 
\end{figure}

Figure 8 presents a combined method of cell division and selection of the optimal number of antennas. Using the analysis of Figure 6, the proposed method is expected to have the best performance among the existing methods. By increasing the maximum transmission power, how to change secure EE is shown. Due to this figure, the combined method has better results than other methods and creates a significant improvement in system performance.

\begin{figure}
	\centering
	\includegraphics[scale=0.6]{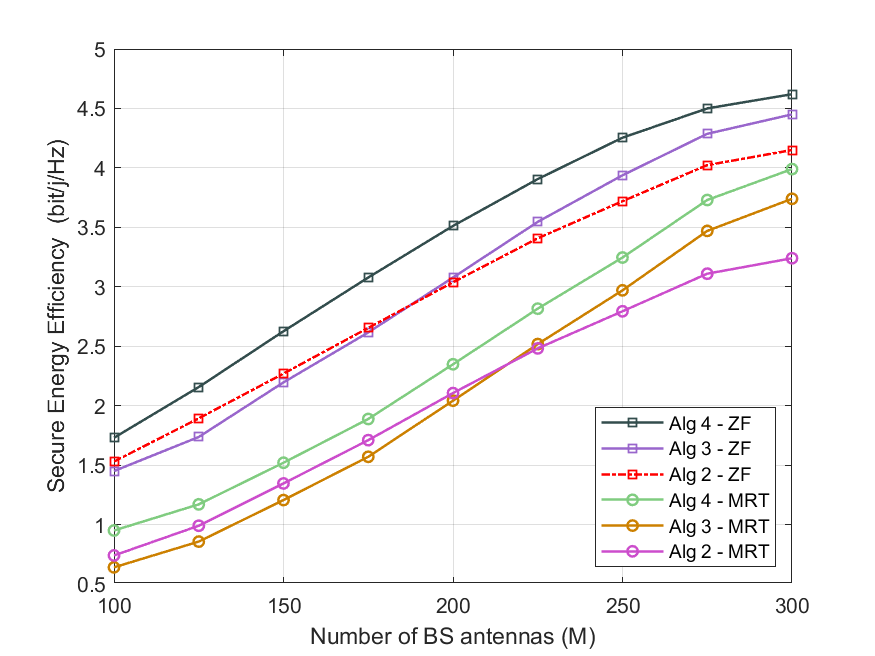}
	\caption{Secure energy efficiency versus the number of transmit antennas.} 
\end{figure}

Finally, figure 9 examines the performance of secure EE by increasing the number of antennas for the proposed joint method. In this figure, it is clearly seen that the combined method has better performance than other methods and improves the system performance.

\begin{figure}
	\centering
	\includegraphics[scale=0.6]{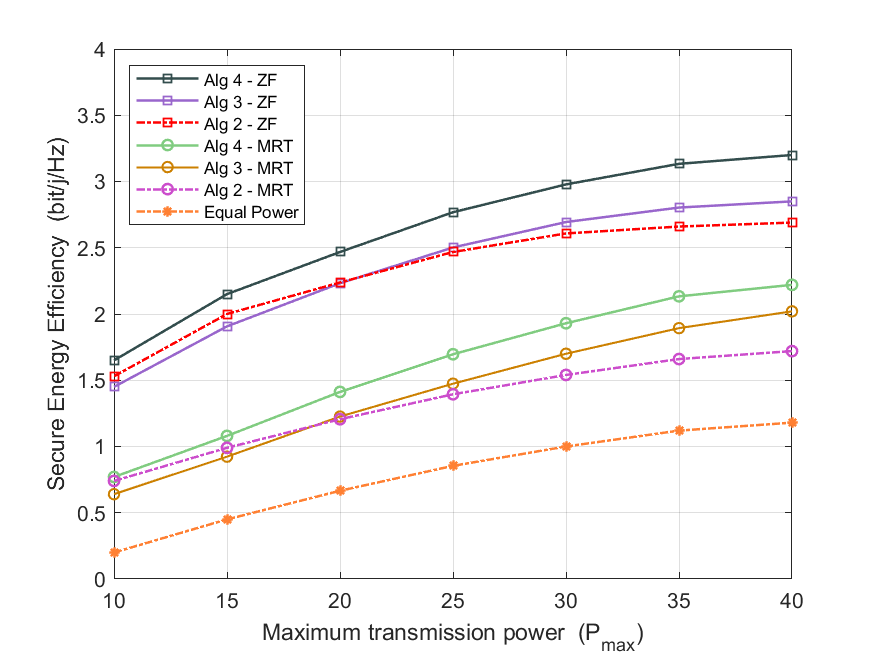}
	\caption{Secure energy efficiency versus the maximum transmission power ($P_{\max}$).} 
\end{figure}

\section {Conclusion and future works}

This paper examines the performance of a massive MIMO system from the secure EE perspective. In this system, an EVA disrupts system performance. The CSI in this paper is considered to be imperfect. To prevent EVA disruption, the authors have proposed two methods of cell division and optimal antennas selection. System performance has been evaluated with both the ZF and the MRT precodings. The optimization problem is also secure EE maximization, which has two constraints, including the maximum transmission power and the minimum secure user rate. Using linear programming, Lagrange dual function method, and lower bound on the secure user rate, the optimization problem is solved.  Four algorithms are proposed based on different scenarios. Simulation results are presented for the system performance analysis. In the obtained numerical results, it is shown that the cell division method has better performance in low transmission power and the number of fewer antennas than the optimal antenna selection method, while in high transmission power and high number of antennas, the optimal antenna selection method outperforms. The best method among the proposed methods is the combined method of cell division and optimal antenna selection, the results of which are better than the other three solutions. Researchers can choose and employ the efficient method based on their model system and demands, for a given number of antennas and maximum transmission power. One of the future works in the continuation of this paper is to consider the number of EVAs more than one and formulate the system model or to consider the CSI related to EVA imperfectly in order to evaluate the system performance in its general case.

\section*{Declarations}
\textbf{Ethical Approval} Not applicable.
$\\$
\textbf{Competing interests} There is no competing interests or personal relationships that could have appeared to influence the work reported in this paper.
$\\$
\textbf{Authors' contributions} All authors have contributed equally.
$\\$
\textbf{Funding} The authors have not received any funding.
$\\$
\textbf{Availability of data and materials} Enquiries about data availability should be directed
to the authors.

\end{document}